# Gamma-Rays and Neutrinos from Dark Matter

F. W. Stecker[a]

[a]Laboratory for High Energy Astrophysics, NASA Goddard Space Flight Center,
Greenbelt, Maryland 20771, U.S.A.

High energy astrophysical $\gamma$-rays and neutrinos can be produced both by the annihilation and possible slow decay of dark matter particles. We discuss the fluxes and spectra of such secondaries produced by dark matter particles in the universe and their observability in competition with other astrophysical $\gamma$-ray signals and with atmospheric neutrinos. To do this, we work within the assumption that the dark matter particles are neutralinos with are the lightest supersymmetric particles (LSPs) predicted by supersymmetry theories.

## 1. THE NATURE OF THE DARK MATTER

The nature of the dark matter is the subject of a number of papers in these proceedings. For our purposes here, it is sufficient to sumarize some relevant points, *i.e.*, with densities expressed as a fraction of the closure density of the universe, $\Omega$, the fraction in baryons, $\Omega_b$ derived from big bang nucleosynthesis arguments, is much less than the total gravitating value of $\Omega$ required by studies of the large scale structure of the universe and by observed large scale dynamics of the universe. This would imply that the bulk of the dark matter in the universe is non-baryonic.

The mixed dark matter model with a total $\Omega = 1$ [1] predicted fluctuations in the cosmic background radiation [2,3] which were found to be in good agreement with the later COBE measurements. The best agreement appears to be found for $\sim 20\%$ hot dark matter, of which massive neutrinos are the most likely candidates, and $\sim 80\%$ cold dark matter [4–8]. The hot dark matter may consist of 2 or 3 neutrino flavors with degenerate masses [1,9,10]. The most popular cold dark matter particle candidates, as discussed elsewhere in these proceedings, are axions and neutralinos. Of these, the neutralinos, the LSPs from supersymmetry (SUSY) theory, produce high-energy $\gamma$-rays and neutrinos as well as other particles such as the antiprotons [11–14], and positrons [13,15] discussed elsewhere in these proceedings.

Neutralinos can produce $\gamma$-rays and neutrinos in two possible ways. The most guaranteed way is by mutual pair annihilations, which can take place because the neutralinos are Majorana fermions. Indeed, it is the bulk of this self-destruction, which occurs in the very early universe, which determines the value for $\Omega_\chi$ (we will hereafter designate the dark matter neutralinos by the letter $\chi$). The second way in which dark matter can produce $\gamma$-rays and neutrinos is if the LSP is allowed to decay to non-supersymmetric, ordinary particles.

Cosmologically important $\chi$ particles must annihilate with a weak cross section, $\langle\sigma v\rangle_A \sim 10^{-26}$ cm$^3$s$^{-1}$; calculations show that such cross sections lead to a value for $\Omega_\chi \sim 1$ with $\Omega_\chi \propto \langle\sigma v\rangle_A^{-1}$. The fact that SUSY neutralinos are predicted to have such weak annihilation cross sections is one reason why they have become popular dark matter candidates (*e.g.* [16]).

In the minimal SUSY model (MSSM), $\chi$ can be generally described as a superposition of two gaugino states and two Higgsino states. Grand unified models with a universal gaugino mass generally favor states where $\chi$ is almost a pure B-ino ($\tilde{B}$) (*e.g.* [17]), but other states such as photinos and Higgsinos are generally allowed by the theory. Indeed, Kane (these proceedings) presents possible accelerator evidence from CDF that $\chi$ may be a Higgsino of mass $\sim 40$ GeV. Theoretical models for particle dark matter are discussed in detail elsewhere in these proceedings.

## 2. ANNIHILATION TO GAMMA-RAYS

There are basically two types of spectra produced by $\chi\chi$ annihilations, *viz.*, (1) $\gamma$-ray continuum spectra from the decay of secondary particles produced in the annihilation process, and (2) $\gamma$-ray lines, produced primarily from the process $\chi\chi \to \gamma\gamma$.

### 2.1. Continuum Gamma-Rays

The continuum $\gamma$-ray production spectrum from $\chi\chi$ annihilation can be calculated for different types of neutralinos using the appropriate branching ratios for annihilation into fermion-antifermion pairs (*e.g.*, for photinos ($\tilde{\gamma}$) B.R. $\propto \beta_f m_f^2 Q_f^4$ and for Higgsinos ($\tilde{h}$) B.R. $\propto \beta_f m_f^2$).

The continuum spectra are generally the result of hadronic cascades leading to the production and decay of neutral pions [13,18,19]; however, the radiative decay $B^* \to B\gamma$ can play a role in hardening the spectrum if the dark matter particles are Higgsinos [19]. One can employ a program such as the Lund Monte Carlo program [20] to follow the quark jet cascades and subsequent particle decays in order to obtain a final $\gamma$-ray production spectrum [19].

The cosmic $\gamma$-ray flux from $\chi\chi$ annihilation is proportional to the line-of-sight integral of the *square* of the $\chi$ particle density times $\langle\sigma v\rangle_A$,

$$\begin{aligned}\phi(E_\gamma) &\simeq 1.8 \times 10^{-6} \langle\ell_8\rangle \rho_{0.3}^2 M_\chi^{-2} \langle\sigma v\rangle_{26} \\ &\times \zeta_\gamma f(E_\gamma) cm^{-2} s^{-1} sr^{-1}\end{aligned} \quad (1)$$

where $\langle\ell_8\rangle$ is the mean line-of-sight integration length through the dark matter galactic halo in units of 8 kpc, $\rho_{0.3}$ is the mass density of the dark matter halo (obtained from galactic rotation curves) in units of 0.3 GeV cm$^{-3}$, $\langle\sigma v\rangle_{26}$ is in units of $10^{-26}$ cm$^3$s$^{-1}$, $\zeta$ is the $\gamma$-ray multiplicity per annihilation and $f(E_\gamma)$ is the normalized energy distribution function of the $\gamma$-rays produced. As can be seen from eq. (1), cosmic $\gamma$-ray fluxes from $\chi\chi$ annihilations are proportional to the annihilation cross section at low energy, which is related to the high temperature annihilation cross section (see, *e.g.*, [16]) and is roughly inversely proportional to $\Omega_\chi$.

Stecker and Tylka [19] discuss in detail the various channels involved in $\gamma$-ray production via $\chi\chi$ annihilation and give the resulting spectra for some lower mass $\chi$ particles. One can generalize from these results to conclude that the $\gamma$-rays from a dark matter halo will probably be unobservable near the galactic plane because of the competing continuum $\gamma$-rays produced by galactic cosmic rays interacting with interstellar gas. Those processes produce much higher fluxes, except perhaps at very high energies if the $\chi$ particle is massive enough.

At high galactic latitudes, the situation is somewhat different, but still unpromising. Here, in addition to the $\gamma$-ray flux produced by cosmic rays in a line-of-sight perpendicular to the galactic disk, there is an extragalactic background flux which is at least as intense as the galactic one at $\sim 0.1$ GeV, and which is expected to dominate over the galactic flux at multi-GeV energies. The extragalactic $\gamma$-ray background flux, as recently estimated [21], is expected to be more substantial in the multi-GeV range than previously thought. This background is expected to be from the jets of unresolved blazars, a type of active galaxy whose relativistically beamed emission is rapidly variable and usually highly polarized. Blazars are the only known source of extragalactic high energy $\gamma$-rays [22]. At least two of these objects, Mrk 421 and Mrk 501, are known to be sources with hard spectra extending to energies above a TeV [23,24]. The latest estimates of the high energy extragalactic $\gamma$-ray background from unresolved blazars (see Figure 1) [21,25], predict fluxes, even in the range between 10 GeV and 1 TeV, which are much larger than the continuum fluxes predicted for $\chi\chi$ annihilation. This dims the hope of observing continuum $\chi$ annihilation $\gamma$-rays from a dark matter halo at high galactic latitudes.

A more promising possibility would be to look for a very hard spectrum of $\chi\chi$ annihilation $\gamma$-rays from a compact dark matter core which might exist at the galactic center [27]. Such a signal would resemble a point source with a $\gamma$-ray detector of angular resolution close to 1° [19]. In fact, a strong source, 2EG J1746-2852, *has been observed* at the galactic center [22]. This source is observed to have a very hard spectrum extending to energies of at least 10 GeV (J. Mattox, private communication). Such a spectrum would be consistent with the hypothesis that the source of the emission is annihilation radiation from a dark matter core [19].

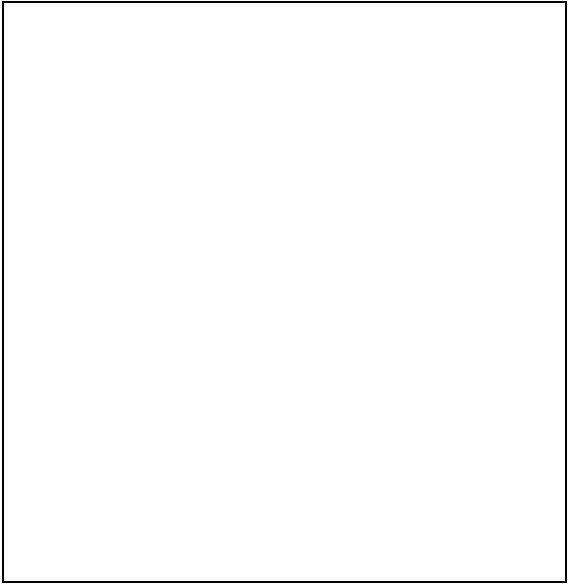

Figure 1. The predicted flux of extragalactic background $\gamma$-rays from unresolved blazars [21,25]. The preliminary data obtained by the EGRET detector on the Compton Gamma Ray Observatory satellite [26] are also shown.

### 2.2. Gamma-Ray Lines

There is also the possibility of observing the two-photon annihilation line from $\chi\chi$ annihilation [28] with a $\gamma$-ray detector of sufficient energy resolution. The general considerations for observability of this line were discussed in detail in Ref. [29].

The energy of the $\chi\chi \to \gamma\gamma$ decay line is $E_\gamma = M_\chi$. The line width is given by Doppler broadening. For galactic halo particles, this width is roughly $\beta_\chi M_\chi \sim 10^{-3} M_\chi$. The line flux, $\phi(M_\chi) \propto \Omega_\chi^{-1}$, as discussed in the previous section.

Therefore, upper and lower limits on $\Omega$ yield lower and upper limits on the $\gamma$-ray line flux respectively. Other limits can also be obtained in flux-energy space [29]. Accelerator determined lower limits on $M_\chi$ give lower limits on the line energy. Lower limits on the mass of the sfermion exchanged in the annihilation process give upper limits on $\langle\sigma v\rangle_A$ since $\langle\sigma v\rangle_A \propto M_{\tilde{f}}^{-4}$. In fact, since the particle density $n_\chi = \rho_\chi/M_\chi$ and $\langle\sigma v\rangle_A \propto M_\chi^2/M_{\tilde{f}}^4$, it follows from eq. (1) that $\phi(E_\gamma) \propto M_{\tilde{f}}^{-4}$. Further limits are obtained from the inequality $M_{\tilde{f}} \geq M_\chi$, which is the tautology following from the condition that $\chi$ be the LSP.

If we assume that annihilations occur mainly through slepton exchange, i.e., $M_{\tilde{q}} \gg M_{\tilde{l}}$, we can obtain an upper limit on the $2\gamma$ line flux. This is because LEP 1.5 gives a lower limit of $\sim 65$ GeV on the slepton mass, whereas the substantially higher squark mass lower limit of $\sim 150$ GeV would imply much lower fluxes, since $\phi_\gamma \propto M_{\tilde{f}}^{-4}$.

Using the latest estimates of the high energy extragalactic $\gamma$-ray background from unresolved blazars [21,25] (see Figure 1), and the latest lower limits on the sfermion masses, one finds that even with a 10% energy resolution, the $\chi$ annihilation line from dark matter in the galactic halo will be difficult to observe above the extragalactic background. However, as in the continuum case discussed in the previous section, it may be possible to observe a $2\gamma$ line from a compact isothermal core [27] at the galactic center with $\sim$10% energy resolution, an angular resolution of $\sim 1°$, and an exposure of 1 m$^2$sr-yr (see discussion in previous section).

### 3. ANNIHILATION TO NEUTRINOS

Neutrinos from $\chi\chi$ annihilation will be produced in galactic and extragalactic space with energies and fluxes similar to the fluxes of the $\gamma$-rays produced. This being the case, and given the much smaller cross sections for neutrino interactions than for $\gamma$-ray interactions, one might presume that one would require an incredibly large neutrino detector, perhaps built on the moon to get above the neutrinos produced by cosmic rays in the Earth's atmosphere, to detect dark mat-ter neutrinos. However, the small neutrino cross section, and the resulting transparency of matter to neutrinos, can also work in favor of dark matter neutrino detection. The Sun can attract and gravitationally focus galactic $\chi$ particles, where they can be scattered by solar protons and captured. They then lose energy by further collisions, eventually finding themselves concentrated in the Sun's core where their annihilation rate is enhanced [30]. The $\chi$ density in the solar core builds up to the point where the $\chi\chi$ annihilation rate reaches equilibrium with the solar $\chi$ capture rate [31]. Whereas the other $\chi$ annihilation products will not escape from the Sun's core, the neutrinos will, providing a potentially detectable point-source signal with a characteristic spectrum $e.g.$ [32]. The neutrinos from $\chi\chi$ annihilations will have typical energies of at least several GeV, perhaps tens of GeV. They are thus easily distinguishable from other solar neutrinos and in the energy range for neutrino telescopes presently planned or under construction.

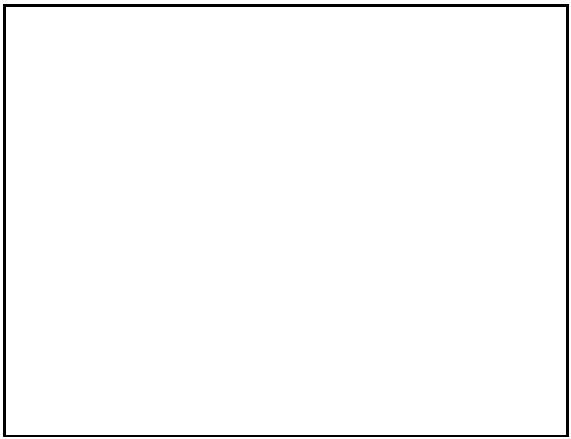

Figure 2. Relative event rates for 15 GeV solar $\chi$ particle annihilaton neutrinos (solid line) and atmospheric cosmic-ray produced neutrinos (dashed line). The $\chi$ neutrino spectrum was computed using the Lund Monte Carlo program as in [14,19].

Since the solar $\chi\chi$ annihilation rate equals the capture rate, resulting in a few high energy neutrinos per annihilation, and since the distance to the Sun is well known, it is easy to make rough estimates of the solar $\chi\chi$ annihilation neutrino flux. This flux should be detectable with a neutrino telescope of suitable detecting area (see, $e.g.$, [33]. The detection event rate $r(E) \propto E^2 \phi(E)$, where $\phi(E)$ is the neutrino flux. This is because the $\sigma_{\nu N} \propto E$ and the range of the detected muon produced by the $\nu N$ interaction is also linear in energy. Thus, detectability is skewed to higher energies by a factor of $E^2$.

Figure 2 shows the relative event rate for cosmic-ray produced atmospheric $\nu$'s and solar $\nu$'s from $\chi\chi$ annihilation calculated for $M_\chi = 15$ GeV. Because of the steepness of the atmospheric $\nu$ spectrum, it should be possible to observe $\nu$'s from annihilations of higher mass $\chi$ particles as well.

## 4. DARK MATTER DECAY

SUSY theories involve a multiplicative quantum number called $R$-parity, which is defined so that it is even for ordinary particles and odd for their SUSY partners. Thus, if R-parity is conserved, as is usually assumed, the LSP is completely stable, making it a potential dark matter candidate. However, such may not be the case. R-parity may be very weakly broken, allowing the LSP to decay with branching ratios involving $\gamma$-rays and neutrinos ($e.g.$ [34]). For $\chi$ particles to be the dark matter, their decay time should be considerably longer than the age of the universe. Of course, invocation of $\chi$ decay involves a higher order of speculation and SUSY model building.

The possible radiative decay $\chi \to \nu + \gamma$ will give a $\gamma$-ray line with energy $E_\gamma = M_\chi/2$. Such a line has the potential of being more intense than the annihilation line. Whereas the $\chi\chi$ annihilation rate and consequent line flux is cosmologically limited by requiring $\Omega_\chi$ to be a significant fraction near 1 (see previous discussion), the decay line flux is limited only by the particular physical SUSY model postulated and constraints from related accelerator and astrophysical data (see $e.g.$, some of the general discussion in [34]).

There is a distinct difference between the expected celestial angular distributions of $\chi\chi$ annihilation $\gamma$-rays and $\chi$ decay $\gamma$-rays. This difference can be used to observationally determine whether multi-GeV $\gamma$-ray lines are from $\chi\chi$ annihilation or $\chi$ decay. Whereas the annihilation flux of dark matter halo $\gamma$-rays is proportional to the line-of-sight integral of $n_\chi^2$, a $\chi$ decay flux would be proportional only to the column density of $\chi$ particles, i.e., the line-of-sight integral of $n_\chi$. Thus, decay photons would come from dark matter decaying throughout the universe and would have a much more isotropic distribution than that expected of annihilation photons. The angular ditribution of annihilation photons from an isothermal dark matter halo exhibits significant anisotropy because of the $n_\chi^2$ factor (e.g., see Ref. [35]).

## 5. CONCLUSION

Cosmic $\gamma$-rays and solar neutrinos produced by SUSY dark matter particles are potentially observable with future telescopes of sufficient sensitivity and energy resolution and with sophisticated techniques for subtracting out events from other sources involving other physical processes. Such observations, in conjunction with accelerator results, may enable us to finally determine the nature of the dark matter.